# Effect of acetylene links on electronic and optical properties of semiconducting graphynes

Yang Li[1†], Junhan Wu[1†], Chunmei Li[1], Qiang Wang[1, 2*], Lei Shen[3*]

[1] *School of Materials and Energy, Southwest University, Chongqing, 400715, China*

[2] *Chongqing Key Laboratory for Advanced Materials and Technologies of Clean Energies, Southwest University, Chongqing 400715, P. R. China*

[3] *Department of Mechanic Engineering & Engineering Science, National University of Singapore, Singapore, 117575, Singapore*

**Abstract:** The family of graphynes, novel two-dimensional semiconductors with various and fascinating chemical and physical properties, has attracted great interest from both science and industry. Currently, the focus of graphynes is on graphdiyne, or graphyne-2. In this work, we systematically study the effect of acetylene, i.e., carbon-carbon triple bond, links on the electronic and optical properties of a series of graphynes (graphyne-*n, where n* = 1-5, the number of acetylene bonds) using the *ab initio* calculations. We find an even-odd pattern, i.e., *n* = 1, 3, 5 and *n* = 2, 4 having different features, which has not be discovered in studying graphyne or graphdyine only. It is found that as the number of acetylene bonds increases, the electron effective mass increases continuously in the low energy range because of the flatter conduction band induced by the longer acetylene links. Meanwhile, longer acetylene links result in larger redshift of the imaginary part of the dielectric function, loss function, and extinction coefficient. In this work, we propose an effective method to tune and manipulate both the electronic and optical properties of graphynes for the applications in optoelectronic devices and photo-chemical catalysis.

# 1  Introduction

The large variety of carbon allotropes, showing different physical and chemical properties, is due to the different carbon hybridizations, i.e, sp, $sp^2$, and $sp^3$. For example, the natural three-dimensional (3D) graphite and diamond are formed through $sp^2$ or $sp^3$ hybridizations of carbon atoms, respectively. Meanwhile, the $sp^2$ hybridization occurs in some novel man-made carbon allotropes, such as fullerene [1], carbon nanotube [2] and graphene [3]. In 1987, the concept of sp-$sp^2$ hybridized graphyne-$n$ was theoretically proposed by Baughman[4], where the $n$ indicated the number of carbon-carbon *triple* (acetylene) bonds in graphyne (see Fig.1). Accordingly, there are several kinds of structures based on the polymerization mode, such as graphyne ($n = 1$), graphdiyne ($n = 2$), graphyne-3 ($n = 3$) and so on. After the successful synthesis of graphyne ($n = 1$) in the experiment, graphyne has been of particular interest to its unique semiconducting electronic structure and extensively applications in many fields, such as catalysis, sensor, transistor, energy storage (see reviews [5,6] and references therein). Simultaneously, engineering of tuning the electronic structure by simply constructing the acetylene bonds $n$ has been attracted more and more attention to this kind of 2D materials theoretically and experimentally.

Recently, 2D semiconducting graphyne-2 has been synthesized on the copper surface by the cross-coupling reaction [7-9]. Soon, this type of 2D materials attracts great attention in many research fields, such as catalysis, energy storage, water purification, and optoelectronic devices, due to its large interlayer distance, unique porous structure, large specific surface area, and high conductivity [10-19]. Theoretical calculations reveal that graphyne-2 has higher electron mobility than graphene [20, 21]. Kuang *et al*., [22] further pointed out that the electron mobility and photoconversion efficiency of perovskite solar cells with graphyne-2 doped was significantly improved, which paves a way for optoelectronic applications of graphyne-2. Wang *et al*., [17] synthesized graphyne-2 composites by the hydrothermal method, which exhibited excellent photocatalytic degradation of the methylene blue. The π-conjugated structure in graphyne-2 makes it efficient to receive photogenerated electrons in the conduction band, and to suppress the recombination of electrons and holes. Luo *et al*., [23] found that the multibody effect had a significant impact on the electronic structure and optical absorption of graphyne-2 in both the theory and experiment. Due to the one more acetylene bond in graphyne-2 compared with graphyne-1, the graphyne-2 has larger porous and much softer character than graphyne-1, which indicates that graphdiyne could easily hybrid with other materials for optical application [5, 6]. The difference of the electronic structures and mechanic properties between graphyne and graphdiyne as well as the resulting different potential applications, has been accelerated the engineering and application of the graphyne-*n* family, especially the properties of graphyne-*n* with longer acetylene links beyond *n* = 1 and 2.

In this paper, the electronic and optical properties of five members in the graphyne family, i.e., graphyne-*n* (*n* = 1-5) are systemically investigated using the *ab initio* calculations. It was found that the length of the acetylene links will greatly change the feature of the energy bands near the Fermi level. Thus, both the electronic and optical properties of this type of 2D materials could be feasibly tuned and manipulated for optoelectronic devices and photo-chemical catalysis applications. This may open a way for exploring the extended graphynes in optoelectronic applications.

## 2   Computational Methods

In this work we carried out *ab initio* calculations with the CASTEP module[24], which was implanted in the framework of the density functional theory (DFT) [25] using the generalized gradient approximation (GGA) in the parameterization of Perdew-Burke-Ernzerhof (PBE) format exchange-correlation functional[26]. The Grimme [27] under dispersion correction (DFT-D) was used to improve the calculation accuracy of the weak interaction in 2D graphynes. The electron-ion interactions were described by the Vanderbilt ultra-soft pseudopotentials (US-PP) [26]. The convergence test and geometric optimization of the graphyne-*n* unit cell were performed firstly. The kinetic cutoff energy used for plane wave expansions was 650 eV. For graphyne-1, graphyne-2 and graphyne-3, the *K* point meshes of $11 \times 11 \times 1$ were used in the first Brillouin-zone with the Monkhorst-Pack [28], while for graphyne-4 and graphyne-5 with large unit cells, the Brillouin zone integrations were performed using a Monkhorst–Pack grid of $8 \times 8 \times 1$. The vacuum layer thickness was set to 15 Å to eliminate the interlayer interaction. Each calculation was converged when the total energy changes during the geometry optimization process were less than $1\times10^{-5}$ eV/atom, and the force per atom and the residual stress of the unit cell was less than 0.01 eV/Å and 0.05 GPa, respectively. The maximum displacement between cycles was less than 0.005 Å when the convergence reached.

## 3 Results and discussion

### 3.1 Geometric structures

The geometric structures of the unit cells of graphyne-1, graphyne-2, graphyne-3, graphyne-4 and graphyne-5 are shown in **Figure 1**, respectively. The size of the cavity of graphynes is proportional to the length of the acetylene linkages. The structural stability of graphynes can be estimated by the cohesive energy, which is defined as follows [29]

$$E_{coh} = \frac{n \times E_{atom} - E_{tot}}{n}, \qquad (1)$$

where $E_{coh}$ is the cohesive energy of graphynes, $n$ is the number of carbon atoms in a unit cell, $E_{atom}$ and $E_{total}$ is the energy of a single carbon atom and the total energy in a unit cell, respectively. The details of the lattice constant, cohesive energy and comparison with other reports are shown in **Table 1**. As can be seen, the calculated lattice parameters in this work are in good agreement with previously reported works. Our cohesive energies are slightly higher than other results, which might be due to the different pseudopotentials used.

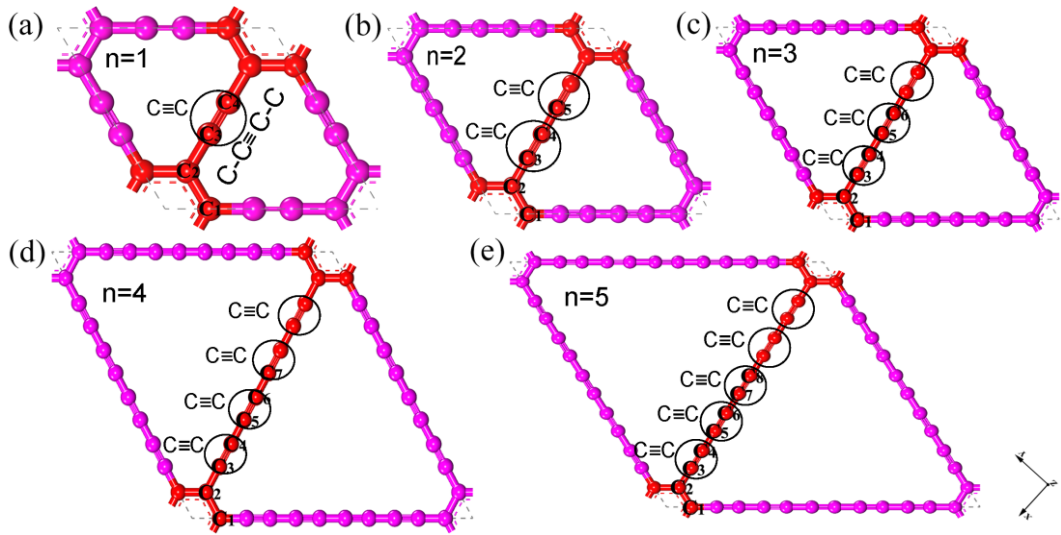

**Figure 1 The optimized geometric structures of unit cells of 2D graphynes**. Each graphyne is named with an index "*n*" which indicates the number of carbon-carbon *triple* bonds in a link, highlighted in red, between two adjacent hexagons. (a) Graphyne-1, (b) Graphyne-2, (c) Graphyne-

3, (d) Graphyne-4 and (e) Graphyne-5.

It is noted that the cohesive energy is the energy required for separating the neutral atoms in the ground state at 0 K [30]. Thus, the larger the $E_{coh}$ is, the more stable the crystal structure is. According to the calculated cohesive energies in **Table 1**, it can be found that the planar two-dimensional structure of graphyne-1 is the most stable. Meanwhile, the cohesive energy of graphynes decreases gradually with the increase of the number of acetylene bonds (*n*).

**Table 1** Lattice constants and cohesive energies of graphynes.

|  |  | Graphyne | Graphdiyne | Graphyne-3 | Graphyne-4 | Graphyne-5 |
|---|---|---|---|---|---|---|
| Lattice constant | This work | 6.872 | 9.436 | 12.011 | 14.576 | 17.592 |
| (Å) | Other works | 6.86[a], 6.877[b], 6.89[c] | 9.44[a], 9.46[c], 9.490[e] | 12.02[a], 12.04[c], 12.43[d] | 14.6[a], 14.60[c] | - |
| Cohesive energy | This work | 8.635 | 8.513 | 8.450 | 8.419 | 8.397 |
| (eV atom$^{-1}$) | Other works | 7.95[a], 7.21[e] | 7.78[a], 7.87[e] | 7.70[a] | 7.66[a] | - |

[a] ref. [31]; [b] ref. [32]; [c] ref. [33]; [d] ref. [34]; [e] ref. [29]

## 3.2 Electronic properties

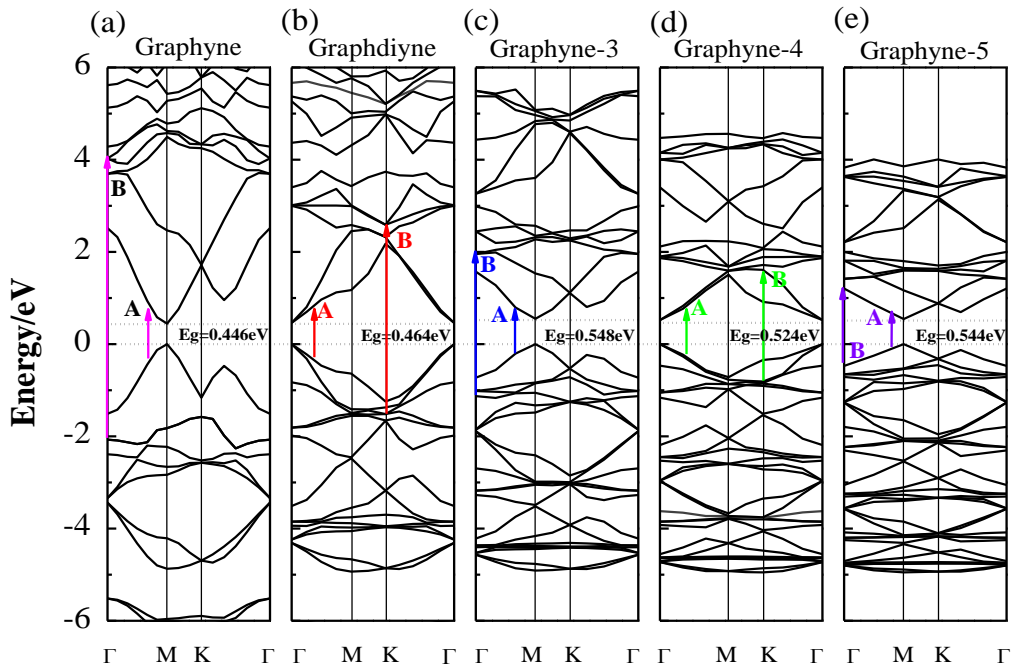

**Figure 2** The energy band structures of graphyne-*n*. $E_g$ is the direct bandgap. *A* and *B* indicate the possible electron

excitation, hopping from the valence band to conduction band.

**Figure 2** shows the band structures of graphynes. It can be seen that all of them are direct bandgap ($E_g$) semiconductors with a bandgap of 0.446, 0.464, 0.548, 0.524 and 0.544 eV, respectively. It is worth noting that the bandgap size is not simply linearly proportional to the number of acetylene bonds ($n$). Interestingly, the position of the direct bandgap of graphyne-$n$ with odd $n$, i.e., $n$ = 1, 3, 5, is located at the M-point, while it is at the Γ-point for even $n$. Meanwhile, the energy-band dispersions at the bottom of the conduction band and the top of valence band are quite similar for all graphynes, which indicates that they have similar effective mass of both electrons and holes.

The effective mass of electrons and holes in semiconductors is an essential parameter, which greatly affect the performance of electronic and/or optical devices. The effective mass of graphynes is calculated by the following equation [35]

$$\frac{1}{m^*} = \frac{1}{\hbar^2}\frac{\partial^2 E(k)}{\partial k^2}. \tag{2}$$

The effective mass in the conduction band ($m_c$, holes) and valence band ($m_v$, electrons) is listed in **Table 2**. It is found that the effective mass is isotropic from M to K and Γ for graphyne-$n$ with the even ($n$ = 2, 4) acetylene bonds, while it is anisotropic if $n$ is odd. Our calculations show that the low energy levels of the conduction band are mainly contributed by the 2p state of carbon atoms, where the electrons have a quite small effective mass under the excitation of photons. This indicates that it facilitates the formation of the effective photogenerated electrons and the transferred charge carriers, while more effective photogenerated holes would take place in the valence-band region.

**Table 2** The effective mass of graphynes in the conduction ($m_c$) and valence band ($m_v$), and band gap ($E_g$).

| Structure | $m_c/m_0$ | | $m_v/m_0$ | | $E_g$ (eV) |
|---|---|---|---|---|---|
| | Γ→M | M←K | Γ→M | M←K | |
| Graphyne-1 | 0.146(0.15[a],0.21[b]) | 0.086(0.063[a],0.087[b]) | 0.150(0.17[a],0.22[b]) | 0.068(0.066[a],0.0901[b]) | 0.446 at M (0.46[c],0.47[d,e]) |
| Graphyne-2 | 0.080(0.073[a]) | | 0.074(0.075[a]) | | 0.464 at Γ (0.48[c],0.46[f,h]) |
| Graphyne-3 | 0.082(0.099[a]) | 0.053(0.081[a]) | 0.106(0.12[a]) | 0.081(0.085[a]) | 0.548 at M (0.56[d]) |
| Graphyne-4 | 0.078(0.081[a]) | | 0.110(0.080[a]) | | 0.524 at Γ (0.54[d]) |
| Graphyne-5 | 0.101 | 0.091 | 0.133 | 0.120 | 0.544 at M |

a: ref. [31]; b: ref. [32]; c: ref. [33]; d: ref. [36]; e: ref. [37]; f: ref. [38]; h: ref. [39]

It is found that all graphynes have covalent bonds from the Mulliken population (MP) analysis (**Table S1**), implying a good structural stability of graphynes. Moreover, the MP analysis shows that the electronic states of graphynes are mainly contributed by the C-2p state, which is consistent with the PDOS analysis in **Figure S1**. The bond lengths and band populations of $C_1$-$C_2$ ($sp^2$-$sp^2$), $C_2$-$C_3$ ($sp^2$-$sp$), $C_4$-$C_5$ ($sp$-$sp$) and $C_6$-$C_7$ ($sp$-$sp$) bonds of graphynes (see in **Figure 1**) remain constant with the increase of the number of acetylene bonds (*n*). While the bond lengths (band populations) of $C_3$-$C_4$, $C_5$-$C_6$ and $C_7$-$C_8$ ($sp\equiv sp$) bonds (**Figure 1**) are enlarged as *n* increases. These alternate bonding characteristics are also illustrated by the charge-density-difference calculations as shown in **Figure S2**. Such alternate -C≡C-C≡C- structure can energetically stabilize atomic carbon chains or rings which have been reported in many carbon allotropes [40, 41].

### 3.3 Optical Properties

The calculated band structures of graphynes in **Figure 2** show that they all are direct band-gap semiconductors, and the values of band gap are close to that of silicon (0.57 eV GGA-PBE[42]). Thus, we next study the optical properties of graphynes and their potential optoelectronic applications. It is well-known that standard GGA functionals like PBE underestimate the band gap. One of major improvements in the band-gap calculation is to use the hybrid functionals, such as HSE06 [43]. However, both GGA-PBE and HSE06 give similar spectra except for an energy shift of graphynes [44], so only GGA-PBE results are presented in this work.

The complex dielectric function $\varepsilon(\omega)$, a significant parameter to determine the polarization effect inside materials, is calculated as:

$$\varepsilon(\omega) = \varepsilon_1(\omega) + i\varepsilon_2(\omega), \tag{3}$$

where the imaginary part of the dielectric function $\varepsilon_2(\omega)$ describes the absorption of light, the real part of the dielectric function $\varepsilon_1(\omega)$ represents the amplitude modulation that's the resonant absorption of the electron transition [45].

(4)

$$\varepsilon_1(\omega) = 1 + \frac{8\pi^2 e^2}{m^2} \cdot \sum_{V,C} \int_{BZ} d^3k \frac{2}{2\pi} \frac{|e \cdot M_{CV}(K)|^2}{[E_C(K) - E_V(K)]} \times \frac{\hbar^3}{[E_C(K) - E_V(K)]^2 - \hbar^2\omega^2}$$

(5)

$$\varepsilon_2(\omega) = \frac{4\pi^2}{m^2\omega^2} \cdot \sum_{V,C} \int_{BZ} d^3k \frac{2}{2\pi} |e \cdot M_{ev}(k)|^2 \times \delta[E_C(K) - E_V(K) - \hbar\omega]$$

$$I(\omega) = \sqrt{2}\omega \left[\sqrt{\varepsilon_1(\omega)^2 - \varepsilon_2(\omega)^2}\right]^{1/2} \tag{6}$$

$$\sigma(\omega) = \frac{\omega}{4\pi}\varepsilon_2(\omega) + i\left[\frac{\omega}{4\pi} - \frac{\omega}{4\pi}\varepsilon_1(\omega)\right] \tag{7}$$

$$n(\omega) = \frac{(n-1)^2 - k^2}{(n+1)^2 + k^2} \tag{8}$$

$$(\omega) = \frac{\varepsilon_2(\omega)}{\varepsilon_1(\omega)^2 + \varepsilon_2(\omega)^2} \tag{9}$$

$$R(\omega) = \left|\frac{\sqrt{\omega}-1}{\sqrt{\omega}+1}\right|^2, \tag{10}$$

where $m$ is the free electron mass, $e$ the free electric charge, $\omega$ the incident photon frequency, $BZ$ the first Brillouin zone, $|e \cdot M_{CV}(K)|$ the momentum transition matrix element, $K$ the inverted lattice vector, $k$ the extinction coefficient, $C$ the conduction band, $V$ the valence band, $E_C(K)$ and $E_V(K)$ the intrinsic level of the conduction and valence bands. Meanwhile, the absorption coefficient $I(\omega)$, conductivity $\sigma(\omega)$, refractive index $n(\omega)$, loss function $L(\omega)$ and reflectivity $R(\omega)$ can all be deduced from $\varepsilon(\omega)$ with the Kramers-Kronig dispersion [46-48].

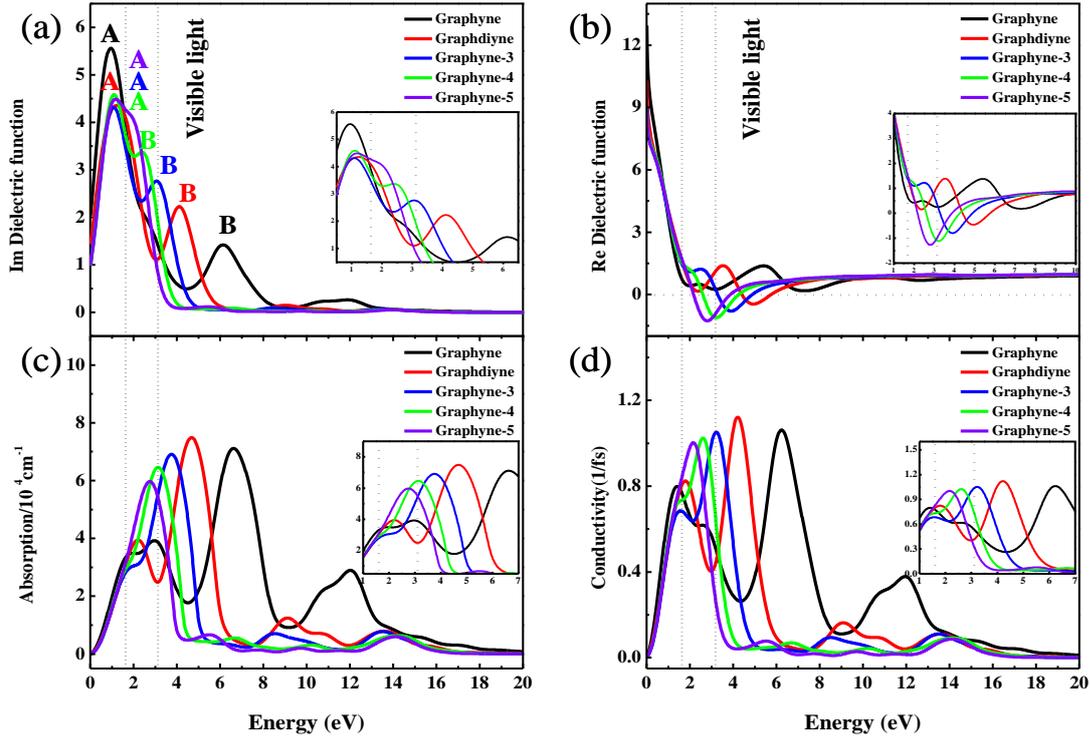

**Figure 3** (a) The imaginary and (b) real parts of complex dielectric function of graphynes. (c) The absorption function and (d) electrical conductivity of graphynes. The visible light region is labeled by two vertical dotted lines. The low-energy part is enlarged in the insets.

**Figure 3** shows the calculated imaginary and real parts of complex dielectric function, absorption function and electrical conductivity of graphynes. The peak position of $\varepsilon_2(\omega)$ is determined by the band gap and the degree of non-localization in the low energy regions. With the increase of the number of acetylene bonds, $\varepsilon_2(\omega)$ of graphynes has redshift as shown in **Figure 3a**. For the real part of the dielectric function $\varepsilon_1(\omega)$, the static dielectric constant (capacitance) is denoted as $\varepsilon_1(0)$ at the zero frequency. The $\varepsilon_1(0)$ of graphynes is strongly dependent to the corresponding band gaps as shown in **Table 3**. Our results in **Figure 3b** show that $\varepsilon_1(\omega)$ dramatically decays to zero with the increase of the photon energies, indicating a resonance of the energy transition between electron and photon in graphynes. **Figure 3b** shows that $\varepsilon_1(\omega)$ of graphyne-2,3,4,5 has negative values (positive for graphyne-1). According to the wave vector equation bellow:

$$\omega^2 \varepsilon_1 = c^2 (\boldsymbol{K} \cdot \boldsymbol{K}), \quad (11)$$

$\varepsilon_1 < 0$ means that the wave vector $\boldsymbol{K}$ is an imaginary number. Furthermore, this negative value region is redshift with the increase of the number of acetylene bonds *(n)* as shown in **Figure3b**.

**Figure 3c** shows that the absorption coefficient in graphynes increases first and consequently decreases as the photon energy goes up. The peak of the absorption coefficient of graphynes shifts to the low energy region and induces a narrow absorption range. Further analysis on the absorption function of graphyne-1 (**Figure 3c**) shows that there are four main absorption peaks at 1.95, 2.96, 6.63 and 12.03 eV, which are well consistent with the experimental values [23]. For graphyne-2 and graphyne-3, the highest peak is located within the ultraviolet region. Thus, they probably may have the applications in ultraviolet protection or detection materials. The shift of the photoconductivity of graphynes with the photon energy is shown in **Figure 3d**. The conductivity peaks of graphyne-1 are at 1.46, 2.59, 6.24 and 11.97 eV, respectively. The profiles of the photoconductivity of graphynes shift to the low-energy region (**Figure 3d**), which is approaching to the energy rage of the visible light.

The peak energy of the dielectric function $\varepsilon_2(\omega)$, energy of electron interband transition, static dielectric constant $\varepsilon_1(0)$ and absorption edges are presented in **Table 3**. For graphyne-1, the primary peaks of $\varepsilon_2(\omega)$ are located at *A* (0.9 eV) and *B* (6.1 eV). Based on the band structure in **Figure 2**, the peak *A* is mainly caused by the transition between the unoccupied states at 0.70 eV and the occupied states at -0.20 eV. Meanwhile, the peak *B* primarily originates from the transition from -2.07 eV to 4.03 eV of the 2p electrons in the valence band of the C atoms. The details of transition of two peaks in $\varepsilon_2(\omega)$ of graphyne-2, graphyne-3 and graphyne-4 are summarized in **Table 3**. For graphyne-5, it is worth noting that the imaginary part of the dielectric function $\varepsilon_2(\omega)$ gives rise to the bimodal which shows just only one peak at 1.1 eV due to the band localization and redshift of the function profile (see in **Figure 3**). Furthermore, the transition between the occupied state at -0.25 eV and the empty state at 0.85 eV corresponds to the peak *A*. Notice that the peak in the imaginary part of the dielectric function $\varepsilon_2(\omega)$ may not correspond to a single inter-band transition (**Figure 2**), other inter-band transitions at the same peak energy may also be possible to occur in the band structure [49, 50].

**Table 3** The energy corresponding to the peak of the imaginary part of the dielectric function $\varepsilon_2(\omega)$, energy of electron inter-band transition, static dielectric constant $\varepsilon_1(0)$ and absorption edges $E_{op}$ of graphynes.

| Structure | Photon energy (eV) | Transition (eV) | $\varepsilon_1(0)$ | $E_{op}$ (eV) |
|---|---|---|---|---|
| Graphyne | A=0.9 | -0.20→0.70 | 12.9 | 5.37 |
| | B=6.1 | -2.07→4.03 | | |
| Graphdiyne | A=1.2 | -0.36→0.84 | 10.3 | 3.57 |
| | B=4.1 | -2.64→1.46 | | |
| Graphyne-3 | A=1.0 | -0.21→0.79 | 8.9 | 2.65 |
| | B=3.0 | -1.10→1.90 | | |
| Graphyne-4 | A=1.1 | -0.27→0.84 | 9.6 | 2.15 |
| | B=2.4 | -0.80→1.60 | | |
| Graphyne-5 | A=1.1 | -0.25→0.85 | 8.9 | 1.79 |

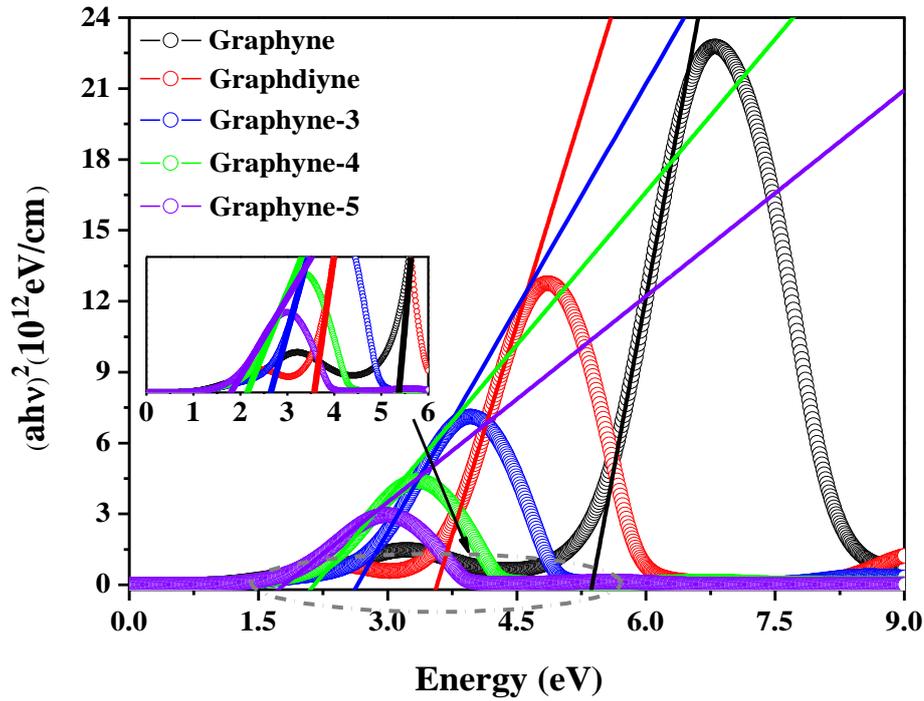

**Figure 4** Optical absorption edges of graphynes, which are enlarged in the inset.

The optical absorption edges of graphynes are shown in **Figure 4**. The optical absorption band edge $E_{op}$ is described by the following extrapolation relationship [51]:

$$\alpha h\nu = A(h\nu - E_{op})^n, \tag{12}$$

where $\alpha$ represents the absorption spectrum, $h\nu$ is the photon energy, $A$ is a function of the refractive index of the material, the reduced mass and the speed of light in vacuum. For direct bandgap semiconductors, $n$ takes 0.5.

The calculated values of the optical absorption band edge in graphyne-$n$ are listed in **Table 3**. It is found that the absorption band edge shifts to the low energy region with the increase of $n$. Moreover, there is a deviation between the absorption band edge and the corresponding band gap. It is mainly due to the electron localization in the free energy level of band structures.

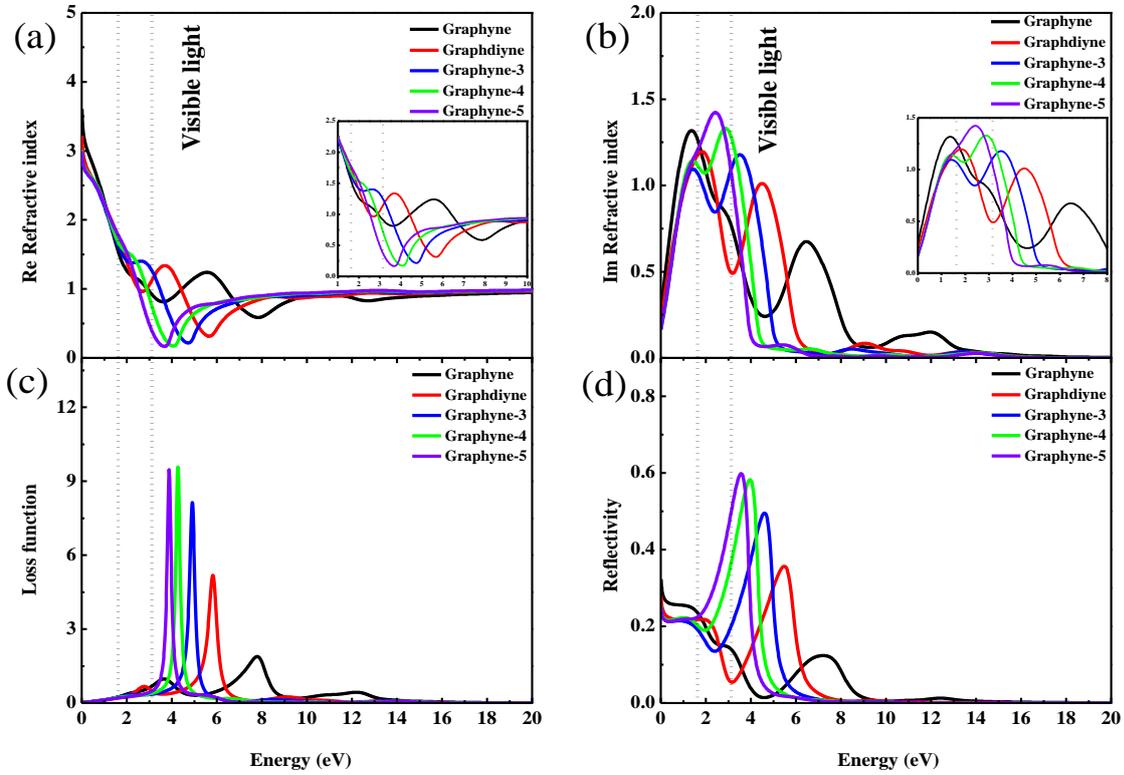

**Figure 5** (a) The real and (b) imaginary parts of complex refractive index of graphynes. (c) The loss function and (d) reflectivity of graphynes. The visible light region is labeled by two vertical dotted lines. The low-energy part is enlarged in the insets of (a) and (b).

The relationship between the complex refractive index $n(\omega)$ and extinction coefficient $k(\omega)$ with the photon energy of graphynes is shown in **Figs. 5a** and **5b**. At the frequency of zero, the static refractive index of graphyne-$n$ ($n$ = 1-5) are located at 3.61, 3.22, 3.00, 3.01 and 2.99 eV, respectively. This indicates that the static index of the refraction strongly depends on the band gap of graphynes. Furthermore, the complex refractive index of graphynes goes down first, and then raises up with the increase of the photon energy. The graphynes have distinct peaks of the energy loss in the calculated profile as shown in **Figure 5c**. They correspond to the region of $\varepsilon_1(\omega)=0$ and $\varepsilon_2(\omega)<0$, which are the resonance peaks of the energy loss function in graphynes. The reflection spectrum of graphynes is shown in **Figure 5d**. Our results show that the reflectivity $R(0)$ of graphyne-1 is approaching to 32.2, and diminishes with the increase of the photon energy. The maximum value is 12.4 % at the photon energy of 7.20 eV. Furthermore, as the acetylene bonds *(n)* increase, the profiles of the energy loss function shift to the low-energy region, and the electron localization and maximum value is enhanced (**Figure S3**).

## 4      Conclusions

In summary, we carried out *ab initio* studies on the geometric, electronic and optical properties of 2D graphyne-*n* sheets, a family of sp-sp$^2$ hybrid materials with acetylene bonds. The odd-even pattern of *n* of graphynes on the position of direct band gaps and the dispersion of the effective mass is revealed by the electronic-structure calculation. Our photoelectron transport results show a redshift of the imaginary part of the dielectric function with the increase of the number of acetylene bonds. Furthermore, the loss function and extinction coefficient moves to the low-energy region with the increase of *n*. These findings show that the optical parameters could be tuned and manipulated by the number of acetylene bonds to fulfill different applications.

**Acknowledgments:** This work was financially supported by by the Sumin Zeng project in Southwest University (ZSM2021008), the Fundamental Research Funds for the Central Universities (XDJK2017B043), and Singapore MOE Tier 1 grant (R- 265-000-691-114).

**Data Availability:** The data that support the findings of this study are available from the corresponding author upon reasonable request.

*Corresponding author wysnu@swu.edu.cn (Qiang Wang)

* Corresponding author shenlei@nus.edu.sg (Lei Shen)

**Supporting Information**

Supporting Information is available free of charge via the Internet at http://pubs.acs.org or from the author.

**Authors Information**

**Corresponding Authors:**

**Qiang Wang -** *School of Materials and Energy and Chongqing Key Laboratory for Advanced Materials and Technologies of Clean Energies, Southwest University, Chongqing 400715, P. R. China; Email: wysnu@swu.edu.cn*

**Lei Shen -** *Department of Mechanic Engineering & Engineering Science, National University of Singapore, Singapore, 117575, Singapore; Email: shenlei@nus.edu.sg*

**Authors:**

Yang Li - *School of Materials and Energy, Southwest University, Chongqing, 400715, China*

Junhan Wu - *School of Materials and Energy, Southwest University, Chongqing, 400715, China*

Chunmei Li - *School of Materials and Energy, Southwest University, Chongqing, 400715, China*

†these authors are equal contribution to this work.

# References:

[1] Krätschmer W.; Lamb Lowell D.; Fostiropoulos K.; Huffman Donald R.. Solid C60: a new form of carbon. Nature **1990**, 347, 354-358.

[2] Iijima, S. Helical microtubules of graphitic carbon. Nature **1991**, 354, 56–58.

[3] Novoselov K. S.; Geim A. K.; Morozov S. V.; Jiang D.; Zhang Y.; Dubonos S. V.; Grigorieva I. V.; Firsov A. A. Electric Field Effect in Atomically Thin Carbon Films. Science **2004**, 36, 666-669.

[4] Baughman R. H.; Eckhardt H..Structure-property predictions for new planar forms of carbon: Layered phases containing Sp2 and sp atoms. J. Chem. Phys. **1987**, 87, 6687-6699.

[5] Yongjun Li, Liang Xu, Huibiao Liu and Yuliang Li. Graphdiyne and graphyne: from theoretical predictions to practical construction. Chemical Society Review **2014**, 43, 2572.

[6] Qing Peng, Albert K Dearden, Jared Crean, Liang Han, Sheng Liu, Xiaodong Wen, Suvranu De. New materials graphyne, graphdiyne, graphone, and graphane: review of properties, synthesis, and application in nanotechnology. Nanotechnology, Science and Applications **2014**, 7, 1-29.

[7] Li G. X.; Li Y. L.; Liu H. B.; Guo Y. B.; Li Y. J.; Zhu D. B.. Architecture of graphdiyne nanoscale films. Chemical Communication **2010**, 46, 3256-3258.

[8] Shang H,; Zou Z.; Li L.; Wang F.; Liu H. B.; Li Y. J.; Li Y. L.. Ultrathin Graphdiyne Nanosheets Grown In Situ on Copper Nanowires and Their Performance as Lithium-Ion Battery Anodes. Angewandte Chemie **2018**, 57, 774-778.

[9] Zhou W. X.; Shen H.; Wu C.; Tu Z.; He F.; Gu Y.; Xue Y.; Zhao Y.; Yi Y.; Li Y.; Li Y.. Direct Synthesis of Crystalline Graphdiyne Analogue Based on Supramolecular Interactions. Journal of the American Chemical Society **2019**, 1, 48-52.

[10] Han Y. Y.; Lu X. L.; Tang S. F.; Yin X. P.; Wei Z. W.; Lu T. B.. Metal-Free 2D/2D Heterojunction of Graphitic Carbon Nitride/Graphdiyne for Improving the Hole Mobility of Graphitic Carbon Nitride. Advanced Energy Materials **2018**, 8, 1702992-1702999.

[11] He J.; Bao K.; Cui W.; Yu J.; Huang C.; Shen X.; Cui Z.; Wang N.. Construction of Large-Area Uniform Graphdiyne Film for High-Performance Lithium-Ion Batteries. Chemistry A European Journal **2018**, 24, 1187-1192.

[12] Liu R.; Liu H.; Li Y.; Yi Y.; Shang X.; Zhang S.; Yu X.; Zhang S.; Cao H.; Zhang G.. Nitrogen-doped graphdiyne as a metal-free catalyst for high-performance oxygen reduction reactions. Nanoscale **2014**, 6, 11336-11343.

[13] Liu R.; Zhou J.; Gao X.; Li J.; Xie Z.; Li Z.; Zhang S.; Tong L.; Zhang J.; Liu Z.. Graphdiyne Filter for Decontaminating Lead-Ion-Polluted Water. Advanced Electronic Materials **2017**, 3, 1700122.

[14] Parvin N.; Jin Q.; Wei Y.; Yu R.; Zheng B.; Huang L.; Zhang Y.; Wang L.; Zhang H.; Gao M.; et al. Few-Layer Graphdiyne Nanosheets Applied for Multiplexed Real-Time DNA Detection. Advanced Materials **2017**, 29, 1606755.

[15] Wang S.; Yi L.; Halpert J. E.; Lai X.; Liu Y.; Cao H.; Yu R.; Wang D.; Li Y.. A Novel and Highly Efficient Photocatalyst Based on P25–Graphdiyne Nanocomposite. Small **2012**, 8, 265-271.

[16] Xue Y.; Zou Z.; Li Y.; Liu H.; Li Y.; Graphdiyne-Supported NiCo2S4 Nanowires: A Highly Active and Stable 3D Bifunctional Electrode Material. Small **2017**, 13, 1700936.

[17] Zhang H.; Xia Y.; Bu H.; Wang X.; Zhang M.; Lou Y.; Zhao M.. Graphdiyne: A promising anode material for lithium ion batteries with high capacity and rate capability. Journal of Applied Physics **2013**, 113, 044309.

[18] Zhang S.; Liu H.; Huang C.; Cui G.; Li Y.. Bulk graphdiyne powder applied for highly efficient lithium storage. Chem. Commun. **2015**, 51, 1834-1837.

[19] Zou Z.; Shang H.; Chen Y.; Li J.; Liu H.; Li Y.; Li Y.. A facile approach for graphdiyne preparation under atmosphere for an advanced battery anode. Chem. Commun. **2017**, 53, 8074-8077.

[20] Padilha J. E.; Fazzio A.; Silva A. J. R.. Directional Control of the Electronic and Transport Properties of Graphynes. The Journal of Physical Chemistry C **2014**, 118, 18793-18798.

[21] Chen J.; Xi J.; Wang D.; Shuai Z.. Carrier Mobility in Graphyne Should Be Even Larger than That in Graphene: A Theoretical Prediction. The Journal of Physical Chemistry Letters **2013**, 4, 1443-1448.

[22] Kuang C.; Tang G.; Jiu T.; Yang H.; Liu H.; Li B.; Luo W.; Luo W.; Li X.; Zhang W.; Lu F.; et al. Highly Efficient Electron Transport Obtained by Doping PCBM with Graphdiyne in Planar-Heterojunction Perovskite Solar Cells. Nano Letters **2015**, 15, 2756-2762.

[23] Luo G.; Qian X.; Liu H.; Qin R.; Zhou J.; Li L.; Gao Z.; Wang E.; Mei W. N.; Lu J.; et al.. Quasiparticle energies and excitonic effects of the two-dimensional carbon allotrope graphdiyne: Theory and experiment. Physical Review B **2011**, 84, 075439.

[24] Perdew J. P.; Burke K.; Ernzerhof M.. Generalized Gradient Approximation Made Simple. Phys. Rev. Lett. **1997**, 78, 1396.

[25] Segall M. D.; Lindan P. J. D.; Probert M.; Pickard C. J.; Hasnip P. J.; Clark S. J.; Payne M. C.. First-principles simulation: ideas, illustrations and the CASTEP code. Journal of Physics: Condensed Matter **2002**, 14, 2717.

[26] Vanderbilt D.. Soft self-consistent pseudopotentials in a generalized eigenvalue formalism. Physical Review B **1990**, 41, 7892.

[27] Grimme S.; Antony J.; Ehrlich S.; Krieg H.. A consistent and accurate ab initio parametrization of density functional dispersion correction (DFT-D) for the 94 elements H-Pu. The Journal of Chemical Physics **2010**, 132, 154104.

[28] Chadi D.J.. Special points for Brillouin-zone integrations. Physical Review B **1977**, 16, 1746.

[29] Puigdollers A. R.; Alonso G.; Gamallo P.. First-principles study of structural, elastic and electronic properties of α-, β- and γ-graphyne. Carbon **2016**, 96, 879-887.

[30] Dickman S., Senozan N. M.; Hunt R. L.. Thermodynamic Properties and the Cohesive Energy of Calcium Ammoniate. The Journal of Chemical Physics **1970**, 52, 2657-2663.

[31] Ducéré, J.M.; C. Lepetit C.; Chauvin R.. Carbo-graphite: Structural, Mechanical, and Electronic Properties. The Journal of Physical Chemistry C **2013**, 117, 21671-21681.

[32] Naritta N.; Nagai S.; Suzuki S.; Nakao K.. Optimized geometries and electronic structures of graphyne and its family. Physical Review B **1998**, 58, 11009.

[33] Yue Q.; Chang S.; Kang J.; Qin S.; Li J.. Mechanical and Electronic Properties of Graphyne and Its Family under Elastic Strain: Theoretical Predictions. The Journal of Physical Chemistry C **2013**, 117, 14804-14811.

[34] Cranford S. W.; Brommer D. B.; Buehler M.. Extended graphynes: simple scaling laws for stiffness, strength and fracture. Nanoscale, **2012**, 4, 7797-7809.

[35] Harald I.; Hans L.; Laszlo M.; David M.. Solid-State Physics: An Introduction to Theory and Experiment. American Journal of Physics **1992**, 60, 1053-1054.

[36] Srinivasu K.; Ghosh S. K.. Graphyne and Graphdiyne: Promising Materials for Nanoelectronics and Energy Storage Applications. The Journal of Physical Chemistry C **2012**, 116, 5951-5956.

[37] Zhou J.; Lv K.; Wang Q.; Chen X. S.; Sun Q.; Jena P.. Electronic structures and bonding of graphyne sheet and its BN analog. The Journal of Chemical Physics **2011**, 134, 174701.

[38] Long M.; Tang L.; Wang D.; Li Y.; Shuai Z.. Electronic Structure and Carrier Mobility in Graphdiyne Sheet and Nanoribbons: Theoretical Predictions. ACS Nano **2011**, 5, 2593-2600.

[39] Zheng Q.; Luo G.; Liu Q.; Quhe R.; Zheng J.; Tang K.; Gao Z.; Nagase S.; Lu J.. Structural and electronic properties of bilayer and trilayer graphdiyne. Nanoscale **2012** 4, 3990-3996.

[40] Shen L.; Zeng M.; Yang S. W.; Zhang C.; Wang X.; Feng Y.. Electron Transport Properties of Atomic Carbon Nanowires between Graphene Electrodes. Journal of the American Chemical Society **2010**, 132, 11481-11486.

[41] Zhang L.; Li H.; Feng Y. P.; Shen L.. Diverse Transport Behaviors in Cyclo[18]carbon-Based Molecular Devices. The Journal of Physical Chemistry Letters 2020, 11, 2611-2617.

[42] Tran F.; Blaha P.. Importance of the Kinetic Energy Density for Band Gap Calculations in Solids with Density Functional Theory. The Journal of Physical Chemistry A **2017**, 121, 3318-3325.

[43] Deák P.; Aradi B.; Frauenheim T.; Janzén E.; Gali A.. Accurate defect levels obtained from the HSE06 range-separated hybrid functional. Physical Review B **2010**, 81, 153203.

[44] Kang J.; Li J.; Wu F.; Li S. S.; Xia J. B.. Elastic, Electronic, and Optical Properties of Two-Dimensional Graphyne Sheet. The Journal of Physical Chemistry C **2011**, 115, 20466-20470.

[45] Okoye C. M. I.. Theoretical study of the electronic structure, chemical bonding and optical properties of KNbO3 in the paraelectric cubic phase. Journal of Physics: Condensed Matter **2003**, 15, 5945.

[46] Kronig R. D. L.. On the Theory of Dispersion of X-Rays. Journal of the Optical Society of America **1926**, 12, 547-557.

[47] Xie Z.; Hui L.; Wang J.; Zhu G.; Chen Z.; Li C.. Electronic and optical properties of monolayer black phosphorus induced by bi-axial strain. Computational Materials Science **2018**, 144, 304-314.

[48] Zheng S.; Wu E.; Feng Z.; Zhang R.; Xie Yuan.; Yu Y.; Zhang R.; Li Q.; Liu J.; Pang W.; et al.. Acoustically enhanced photodetection by a black phosphorus–MoS2 van der Waals heterojunction p–n diode. Nanoscale **2018**, 10, 10148-10153.

[49] Almeida J. S.; Ahuja R.. Tuning the structural, electronic, and optical properties of BexZn1−xTe alloys. Applied Physics Letters **2006**, 89, 061913.

[50] Ma T. H.; Yang C. H.; Xie Y.; Sun L.; Lv W. Q.; Wang R.; Zhu C. Q.; Wang M.. Electronic and optical properties of orthorhombic LiInS2 and LiInSe2: A density functional theory investigation. Computational Materials Science **2009**, 47, 99-105.

[51] Srikant V. Clarke D. R.. Optical absorption edge of ZnO thin films: The effect of substrate. Journal of Applied Physics **1997**, 81, 6357.

## Abstract Graphics

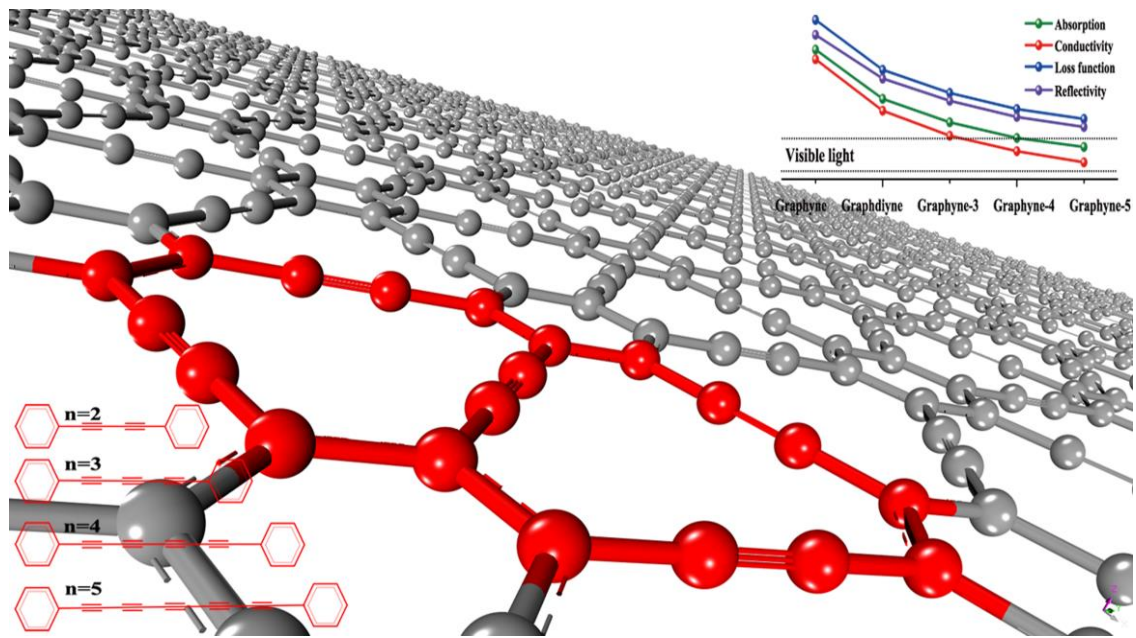